# The Behavior of the Electromagnetic Field Near the Edge of a Resistive Half-Plane

Igor M. Braver, Pinchos Sh. Fridberg, Khona L. Garb, and Iosif M. Yakover

*Abstract*–The present work is aimed at defining the behavior of the electromagnetic field near the edge of a resistive half-plane, taken separately, as well as in conjunction with a perfectly conducting half-plane. The efficiency of accounting for the established field behavior while finding the solution of boundary problems of electrodynamics is illustrated by computational results.

## I. INTRODUCTION

In formulating the boundary value problems of electrodynamics describing structures with sharp edges, account should be taken of the so-called edge condition. This condition requires that finite energy be concentrated near the edge. Given this limitation each of the components of the electric field $\vec{E}$ and magnetic field $\vec{H}$ should be quadratically integrable over any domain containing the edge. More specific knowledge of the behavior of $\vec{E}$, $\vec{H}$ is not necessary for formulating the problem. However, such knowledge can substantially improve the efficiency of the computation of the solution. The latter consideration accounts for numerous works dealing with this subject. To date, we know the behavior of the field near the edge of perfectly conducting and magnetodielectric wedges [1]–[4]. The present work investigates the behavior of $\vec{E}$, $\vec{H}$ near the edge of a resistive half-plane (Fig. 1) on the surface of which two-sided impedance type boundary conditions are fulfilled

$$E_\rho^+ = E_\rho^- = W\left(H_u^+ - H_u^-\right), \quad E_u^+ = E_u^- = -W\left(H_\rho^+ - H_\rho^-\right), \tag{1}$$

where $W$ is surface impedance, and " + " and " – " denote a field at $\varphi = 0$ and $\varphi = 2\pi$, respectively.

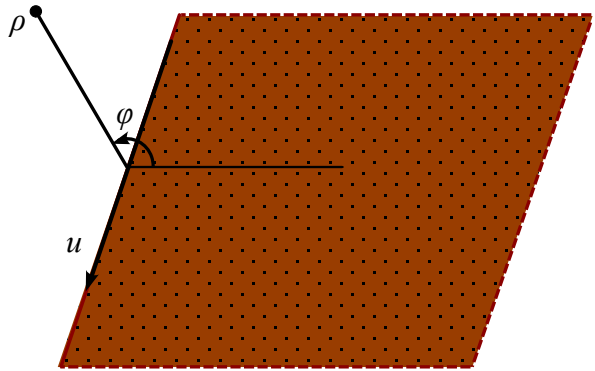

Fig. 1. Resistive half-plane.

While studying the behavior of the field at $\rho \to 0$ we can disregard the dependence of $\vec{E}$, $\vec{H}$ on the $u$ coordinate and consider two independent problems of defining $E_u$, $H_\rho$, $H_\varphi$ ($E$-polarization) and $H_u$, $E_\rho$, $E_\varphi$ ($H$-polarization). Dependence of the fields on time $t$ is chosen in the form of $\exp(-i\omega t)$.

## II. BIFURCATION OF A PLANE WAVEGUIDE BY A RESISTIVE HALF-PLANE

One of the ways to determine the field near the edge is to employ analysis of a problem, having analytical solution. Monograph [5] shows how Meixner's result for the perfectly conducting half-plane can be obtained by analysis of the solution of the problem dealing with bifurcation of a parallel plate waveguide. The solution of a similar problem for a resistive half-plane allows one to determine the field behavior near its edge.

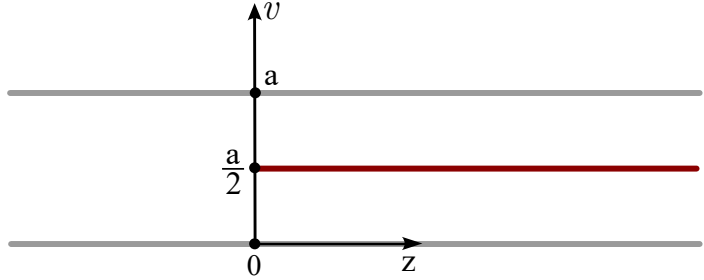

Fig. 2. Plane waveguide divided by a resistive half-plane.

It is assumed that in the region $z = -\infty$ of the parallel plate waveguide (Fig. 2), there is incident either an $LE_{01}$ mode ($s=1$) or an $LM_{01}$ mode ($s=2$). The incident field induces on the resistive half-plane a surface current, the density $\vec{j}$ of which will be expressed in the terms of a dimensionless function $g_s(z)$:

$$j_u(z) = i\frac{V}{\zeta}\frac{k_1}{k}g_1(z), \quad j_z(z) = i\frac{k_1 a}{\pi}g_2(z), \qquad (2)$$

where axis $u$ is perpendicular to Fig. 2; $V$ (or $I$) is the value of the corresponding component of the electric (or magnetic) field of the incident mode at $v = \frac{a}{2}$, $z = 0$;

$k_m = \sqrt{k^2 - \left((2m-1)\frac{\pi}{a}\right)^2}$, $m = 1, 2, 3, \ldots$; $k$ and $\zeta = 120\pi\,\Omega$ are the wavenumber and the wave impedance in free space.

Function $g_s(z)$ is to be sought in the form

$$g_s(z) = \sum_{\nu=1}^{\infty} A_\nu^{(s)} \exp(ih_\nu^{(s)} z). \qquad (3)$$

Here $h_\nu^{(s)}$ are propagation constants of $LE_{0,2\nu-1}$ modes ($s=1$) or $LM_{0,2\nu-1}$ modes ($s=2$) in the waveguide with resistive film. The unknown coefficients $A_\nu^{(s)}$ are determined in analytic form by the technique described in [6]:

$$A_\nu^{(s)} = \left(k_1 - h_\nu^{(s)}\right) a \prod_{m=1}^{\infty} \left(\frac{h_m^{(s)} - k_1}{h_m^{(s)} - h_\nu^{(s)}}\right)_{(m\neq\nu)} \left(\frac{k_m - h_\nu^{(s)}}{k_m - k_1}\right)_{(m\neq 1)}, \qquad (4)$$

where the index ($m \neq \nu$) indicates that at $m = \nu$ the corresponding factor must be assumed equal to 1. The behavior of function $g_s(z)$, and hence current $\vec{j}$, at $z \to +0$ is determined by the dependence of the $A_\nu^{(s)}$ coefficients on $\nu$ as $\nu \to \infty$. To find the nature of this dependence let us first determine an asymptotic expression for $h_\nu^{(s)}$ at high $\nu$. The propagation constants $h_\nu^{(s)} = \sqrt{k^2 - \left(\kappa_\nu^{(s)}\right)^2}$ are expressed in the terms of the transverse wavenumbers $\kappa_\nu^{(s)}$, ordered according to increasing $\operatorname{Re} \kappa_\nu^{(s)}$. From the dispersion equations for $\kappa_\nu^{(s)}$ obtained in [7], we get

$$h_\nu^{(1)} - k_\nu \simeq -\frac{\zeta}{W}\frac{k}{2\pi\nu}, \quad h_\nu^{(2)} - k \simeq -i\frac{\pi}{a}, \quad \nu \to \infty. \tag{5}$$

At $s = 1$ it is convenient for further analysis to transform expression (4) so that

$$A_\nu^{(1)} = \left(k_1 - h_\nu^{(1)}\right) a \left[\frac{\left(h_1^{(1)} - k_1\right)\left(k_\nu - h_\nu^{(1)}\right)}{\left(h_1^{(1)} - h_\nu^{(1)}\right)\left(k_\nu - k_1\right)}\right]_{(\nu \neq 1)} \prod_{m=2}^{\infty} \left\{1 - \frac{\left(k_1 - h_\nu^{(1)}\right)\left(k_m - h_m^{(1)}\right)}{\left(h_m^{(1)} - h_\nu^{(1)}\right)\left(k_m - k_1\right)}\right\}_{(m \neq \nu)}. \tag{6}$$

Taking into account (5) it is easy to see that as $\nu \to \infty$ the infinite product in (6) remains finite and, hence, the decrease of coefficients $A_\nu^{(1)}$ is determined by the factor in square brackets. This allows $A_\nu^{(1)} \sim \nu^{-2}$ as $\nu \to \infty$. For the current density on the film we have

$$j_u(z) \sim \sum_{\nu=1}^{\infty} \frac{1}{\nu^2} \exp\left(-2\pi\nu\frac{z}{a}\right) \tag{7}$$

and having summed with respect to $\nu$ [8], we find

$$j_u(z) \sim D_1 + D_2 z \ln\left(\frac{z}{a}\right) + O\left(\frac{z}{a}\right), \quad z \to +0, \tag{8}$$

where $D_1$ and $D_2$ are constants, independent of $z$.

For $s = 2$ the analysis of expression (4) appears to be more complex. However, there is no need to carry out this analysis, since neither formula (4) nor relation (5) for $h_\nu^{(2)}$ change with replacement of the resistive half-plane with a perfectly conducting one. This allows to conclude that on the resistive half-plane for the current component perpendicular to the edge, Meixner's relation $j_z(z) \sim \sqrt{z}$, $z \to +0$, remains valid.

Thus, having considered two particular problems of diffraction by a resistive half-plane, we have determined the behavior of the current components near the edge. Analyzing these solutions one could also obtain the distribution of all components of the electromagnetic field. However, a more detailed study of the field in the vicinity of the edge will be carried out by a general method, not connected with the solution of any specific problems.

## III. FIELD EXPANSION NEAR THE EDGE OF A RESISTIVE HALF-PLANE (FIG. 1)

Our further aim is to find the distribution of the fields $\vec{E}, \vec{H}$, satisfying (in the source-free region) Maxwell's homogeneous equations and boundary conditions (1). In both *E*- and *H*-polarization, the components to be found are expressed as their scalar function $\psi$ ($\psi \equiv E_u$ or $\psi \equiv H_u$) which is the solution of the two-dimensional Helmholtz equation

$$\frac{1}{r}\frac{\partial}{\partial r}\left(r\frac{\partial \psi}{\partial r}\right) + \frac{1}{r^2}\frac{\partial^2 \psi}{\partial \varphi^2} + \psi = 0, \tag{9}$$

where $r = k\rho$, and it is assumed that $k \neq 0$.

If we were to apply Meixner's technique [1] to find $\psi$ as a power series

$$\psi = r^\tau \sum_{m=0}^{\infty} C_m(\varphi) r^m, \quad \tau \geq 0, \tag{10}$$

we could find only a trivial solution, having the form

$$E_u = \sum_{m=1}^{\infty} Q_m J_m(r) \sin m\varphi, \quad H_\rho = -\frac{i}{\zeta r}\frac{\partial E_u}{\partial \varphi}, \quad H_\varphi = \frac{i}{\zeta}\frac{\partial E_u}{\partial r}; \tag{11a}$$

$$H_u = \sum_{m=0}^{\infty} P_m J_m(r) \cos m\varphi, \quad E_\rho = \frac{i\zeta}{r}\frac{\partial H_u}{\partial \varphi}, \quad E_\varphi = -i\zeta\frac{\partial H_u}{\partial r}, \tag{11b}$$

where $Q_m$, $P_m$ are arbitrary constants, and $J_m(x)$ is a Bessel function. Expressions (11) correspond to a field configuration where $E_u^\pm = E_\rho^\pm = 0$, $H_u^+ = H_u^-$, $H_\rho^+ = H_\rho^-$, and, hence, boundary conditions (1) are fulfilled at any $W$. In other words, the presence of the resistive half-plane does not distort the field distribution (11), because tangential components of $\vec{E}$ on its surface vanish. Of major interest is the distribution of the field which induces on resistive half-plane the electric current with surface density $j_u = -\left(H_\rho^+ - H_\rho^-\right)$, $j_\rho = H_u^+ - H_u^-$. It turned out that such a field is described by a function which cannot be expanded in a power series. Hence, to obtain a nontrivial solution we must alter (10).

A series describing the general solution of an ordinary differential equation is known [9] to contain, besides powers of $r$, powers of $\ln r$ also. Furthermore, the logarithmic function is contained in the current expansion (8). Given all this, we can attempt to supplement expansion (10) with terms proportional to $\ln^n r$. A similar series was employed in [10] to solve the set of differential equations for the longitudinal field components in quadrupole-magnetized ferrite. In our problem the general solution of (9) with conditions (1) can be found in the form

$$\psi = r^\tau \sum_{m=0}^{\infty} \sum_{n=0}^{m} C_{mn}(\varphi) r^m \ln^n r, \quad \tau \geq 0. \tag{12}$$

The requirement $\tau \geq 0$ assures quadratic integrability (over the area with surface element $\rho\, d\rho\, d\varphi$) of the transverse field components in the vicinity of the edge of the half-plane. From this point of view $\tau > 0$ are acceptable at any $C_{mn}$, and $\tau = 0$ only if $C_{00}' = 0$, where prime means differentiation with respect to $\varphi$. Substitution of (12) in (9) leads to an infinite set of differential equations

$$C_{mn}'' + (m+\tau)^2 C_{mn} + 2(m+\tau)(n+1)C_{m,n+1} + (n+2)(n+1)C_{m,n+2} + C_{m-2,n} = 0. \tag{13}$$

From here on $C_{mn} = 0$, if $m < 0$ or $m < n$. Constants of integration arising in solving (13) and the value $\tau$, which ensures nonzero solution are to be determined from boundary conditions (1). Further investigation will be carried out for two separate types of polarization.

## IV. *E*-POLARIZATION

The boundary conditions have the form

$$C_{mn}(0) = C_{mn}(2\pi), \quad C_{mn}'(0) - C_{mn}'(2\pi) = -\frac{i}{w} C_{m-1,n}(0), \tag{14}$$

where $w = \dfrac{W}{\zeta}$. For the dominant coefficient $C_{00}$ there is a homogeneous equation

$$C_{00}'' + \tau^2 C_{00} = 0. \tag{15}$$

Substitution of its general solution $C_{00} = p_{00} \cos \tau\varphi + q_{00} \sin \tau\varphi$, where $p_{00}$, $q_{00}$ are constants, into boundary conditions $C_{00}(0) = C_{00}(2\pi)$, $C_{00}'(0) = C_{00}'(2\pi)$ will produce the set of equations

$$\begin{cases} p_{00}(1 - \cos 2\pi\tau) - q_{00} \sin 2\pi\tau = 0, \\ p_{00}\tau \sin 2\pi\tau + q_{00}\tau(1 - \cos 2\pi\tau) = 0. \end{cases} \tag{16}$$

Nonzero coefficient $C_{00}$ can only exist if the main determinant of the set (16) is zero, which gives us a characteristic equation for $\tau$:

$$\tau \sin^2 \pi\tau = 0. \tag{17}$$

The lowest admissible $\tau$, satisfying (17), is $\tau = 0$. Hence $C_{00} = p_{00}$ is an arbitrary constant and component $E_u$ on the half-plane edge turns out to be finite. The expression for $C_{00}$ is not yet sufficient to determine the behavior of $H_\rho$ and $H_\varphi$. Hence, the next coefficients in expansion (12) must be found.

Let us consider the algorithm for solution of the infinite set (13), (14) at $\tau = 0$. For each fixed *m* we are first of all to find coefficient $C_{mm}$, the differential equation for which is homogeneous and has the solution

$$C_{mm} = p_{mm} \cos m\varphi + q_{mm} \sin m\varphi. \tag{18}$$

Since *m* is an integer, (18) will satisfy homogeneous boundary conditions $C_{mm}(0) = C_{mm}(2\pi)$, $C_{mm}'(0) = C_{mm}'(2\pi)$ at arbitrary constants $p_{mm}$, $q_{mm}$. Let us now sequentially reduce index *n*. The differential equations for coefficients $C_{m,m-1}$ are inhomogeneous and their solution can be written as

$$C_{m,m-1} = p_{m,m-1} \cos m\varphi + q_{m,m-1} \sin m\varphi - p_{mm} m\varphi \sin m\varphi + q_{mm} m\varphi \cos m\varphi. \tag{19}$$

Substitution of (19) into (14), allows one to find

$$q_{mm} = 0, \quad p_{mm} = -\frac{i}{2\pi m^2 w} p_{m-1,m-1}, \tag{20}$$

which immediately determines the general expression for the diagonal coefficients

$$C_{mm} = \frac{p_{00}}{(2\pi i w)^m (m!)^2} \cos m\varphi. \tag{21}$$

It is easy to see that at fixed $m$ from conditions (14) we obtain sets of linear equations for constants $(p_{mm}, q_{mm}); \ldots; (p_{m1}, q_{m1})$. Sequential solution of these sets permits expression of all $(p_{mn}, q_{mn})$ in terms of $(p_{m0}, q_{m0})$. The sought-for function $\psi$ can then be transformed to

$$\psi = \sum_{\mu=0}^{\infty} \{p_{\mu 0} g_\mu(r, \varphi) + q_{\mu 0} J_\mu(r) \sin \mu\varphi\}. \tag{22}$$

The functions $g_\mu$ depend on parameter $w$, contain no undefined constants, and possess asymptotic behavior $g_\mu = O(r^\mu)$ as $r \to 0$. Expression (22), which contains the above-described trivial solution (11a), is the general solution of the *E*-polarization problem. Note, that the next solutions $\tau = 1, 2, 3, \ldots$ of the characteristic equation (17) do not bring about any new functions in (22), but only restrict the lowest value of index $\mu$ to the value equal to $\tau$.

The arbitrary value of constants $p_{\mu 0}$, $q_{\mu 0}$ in (22) is explained by the fact that external sources in the problem under consideration are not specified, hence the problem has an infinite set of solutions. Specification of such sources will determine in each specific case $p_{\mu 0}$, $q_{\mu 0}$ and thus the solution satisfying the uniqueness theorem. Since we are interested in the behavior of the field as $r \to 0$, we shall write only a partial solution, proportional to $p_{00} = U_0$:

$$\begin{aligned} E_u &= U_0 \left[ 1 - \frac{i \cos \varphi}{2\pi w} r \ln r + \frac{i\varphi \sin \varphi}{2\pi w} r - \frac{\cos 2\varphi}{16\pi^2 w^2} r^2 \ln^2 r \right. \\ &\quad \left. + \frac{\cos 2\varphi + 2(\varphi - \pi)\sin 2\varphi}{16\pi^2 w^2} r^2 \ln r + \cdots \right], \\ H_\rho &= -\frac{i}{\zeta} U_0 \left[ \frac{i \sin \varphi}{2\pi w} \ln r + \frac{i(\sin \varphi + \varphi \cos \varphi)}{2\pi w} + \frac{\sin 2\varphi}{8\pi^2 w^2} r \ln^2 r \right. \\ &\quad \left. + \frac{(\varphi - \pi) \cos 2\varphi}{4\pi^2 w^2} r \ln r + \cdots \right], \\ H_\varphi &= \frac{i}{\zeta} U_0 \left[ -\frac{i \cos \varphi}{2\pi w} \ln r + \frac{i(\varphi \sin \varphi - \cos \varphi)}{2\pi w} - \frac{\cos 2\varphi}{8\pi^2 w^2} r \ln^2 r \right. \\ &\quad \left. + \frac{(\varphi - \pi) \sin 2\varphi}{4\pi^2 w^2} r \ln r + \cdots \right]. \end{aligned} \tag{23}$$

Expressions (23) show that unlike component $E_u$, components $H_\rho$ and $H_\varphi$ as $r \to 0$ have logarithmic singularity.

The solution found can be represented by a power series in $\frac{1}{w}$, and allows a term by term limiting transition as $w \to \infty$ (half-plane is absent). In this case all terms containing logarithmic functions disappear in the field expansion. To obtain from (23) another limiting case $w \to 0$ (perfectly conducting half-plane) is rather difficult, since the first terms of this expansion cease to dominate for $w \ll r |\ln r| \ll 1$.

## V. *H*-POLARIZATION

The boundary conditions have the form

$$C_{mn}'(0) = C_{mn}'(2\pi) = -iw\left[C_{m-1,n}(0) - C_{m-1,n}(2\pi)\right]. \tag{24}$$

The solution of the homogeneous differential equation for the main coefficient $C_{00}$, and the use of $C_{00}'(0) = C_{00}'(2\pi) = 0$, will bring us to the set of linear equations for $(p_{00}, q_{00})$:

$$\begin{cases} q_{00}\tau = 0, \\ q_{00}\tau \cos 2\pi\tau - p_{00}\tau \sin 2\pi\tau = 0. \end{cases} \tag{25}$$

Hence, the characteristic equation to determine $\tau$ is

$$\tau^2 \sin 2\pi\tau = 0. \tag{26}$$

From the lowest admissible root of this equation $\tau = 0$ we get the trivial solution (11b). The next admissible value $\tau = \frac{1}{2}$ gives $C_{00} = p_{00} \cos\frac{\varphi}{2}$. The obtained expression $\psi = p_{00}\sqrt{r}\cos\frac{\varphi}{2} + \cdots$ will determine the behavior of the field as $r \to 0$, if $p_{00} \neq 0$. Since $C_{00}$ according to (13) and (24) appears in the equations, which determine subsequent coefficients $C_{mn}$, it is necessary to make sure that the condition for the existence of the latter does not bring about the equality $p_{00} = 0$. Let us thus consider the algorithm for the solution of the infinite set (13), (24) for $\tau = \frac{1}{2}$.

The expressions for the diagonal coefficients

$$C_{mm} = p_{mm}\cos\left(m + \frac{1}{2}\right)\varphi + q_{mm}\sin\left(m + \frac{1}{2}\right)\varphi \tag{27}$$

satisfy the boundary conditions $C_{mn}'(0) = C_{mn}'(2\pi) = 0$ at $q_{mm} = 0$ and any $p_{mm}$. For $n = m - 1$ we find

$$C_{m,m-1} = p_{m,m-1}\cos\left(m + \frac{1}{2}\right)\varphi + q_{m,m-1}\sin\left(m + \frac{1}{2}\right)\varphi - p_{mm}m\varphi\sin\left(m + \frac{1}{2}\right)\varphi \tag{28}$$

and from the conditions (24) we have

$$p_{mm} = \frac{1}{\pi m}q_{m,m-1} = -\frac{4iw}{\pi m(2m+1)}p_{m-1,m-1}. \tag{29}$$

Using (29), for $C_{mm}$ we can write

$$C_{mm} = \left(-\frac{4iw}{\pi}\right)^m \frac{p_{00}}{m!(2m+1)!!}\cos\left(m + \frac{1}{2}\right)\varphi. \tag{30}$$

Further lowering of index $n$ leads to the sets of linear equations for sequential determination of the following pairs of constants: $(p_{m,m-1}, q_{m,m-2});\ldots;(p_{m1}, q_{m0})$. Thus, the boundary conditions do not put limitations on the values of constants $p_{m0}$. The latter, as in case of *E*-polarization, can be determined only after we choose the external sources. Here we give the solution proportional to $p_{m0} = I_0$:

$$H_u = I_0 \sqrt{r} \left[ \cos\frac{\varphi}{2} - \frac{4iw}{3\pi} \cos\frac{3\varphi}{2} r \ln r + \frac{4iw}{3\pi}(\varphi - \pi) \sin\frac{3\varphi}{2} r \right.$$
$$\left. - \frac{8w^2}{15\pi^2} \cos\frac{5\varphi}{2} r^2 \ln^2 r + \cdots \right],$$

$$E_\rho = i\zeta I_0 \frac{1}{\sqrt{r}} \left[ -\frac{1}{2}\sin\frac{\varphi}{2} + \frac{2iw}{\pi} \sin\frac{3\varphi}{2} r \ln r \right.$$
$$\left. + \frac{2iw}{\pi}\left( \frac{2}{3}\sin\frac{3\varphi}{2} + (\varphi - \pi)\cos\frac{3\varphi}{2} \right) r + \frac{4w^2}{3\pi^2} \sin\frac{5\varphi}{2} r^2 \ln^2 r + \cdots \right], \quad (31)$$

$$E_\varphi = -i\zeta I_0 \frac{1}{\sqrt{r}} \left[ \frac{1}{2}\cos\frac{\varphi}{2} - \frac{2iw}{\pi} \cos\frac{3\varphi}{2} r \ln r \right.$$
$$\left. + \frac{2iw}{\pi}\left( -\frac{2}{3}\cos\frac{3\varphi}{2} + (\varphi - \pi)\sin\frac{3\varphi}{2} \right) r - \frac{4w^2}{3\pi^2} \cos\frac{5\varphi}{2} r^2 \ln^2 r + \cdots \right].$$

As we see, in (31) unlike in (23) the term by term limiting transition $w \to 0$ is possible. Terms independent of $w$ describe the behavior of the field at the edge of a perfectly conducting half-plane. In the contrary limiting case $w \to \infty$ the finite number of terms of the solution (31) cannot be employed to describe the field in the region $\frac{1}{w} \ll r|\ln r| \ll 1$.

To illustrate the results obtained while considering the two polarizations we provide Table I, where the sign "$\perp$" marks vectors perpendicular to the edge of the half-plane. The Meixner's result [1] is also given there for comparison. The behavior given in Table I is characteristic of the component $H_u$ not of the full field, but of the difference between the full field and the trivial solution (11b). Any boundary value problem of electrodynamics can be reduced to obtaining this difference. It is noteworthy that the obtained nontrivial solution (31) and trivial solution (11b) have contrary properties of field symmetry about the plane $\varphi = 0, \pi$.

TABLE I

BEHAVIOR OF THE FIELD NEAR THE EDGE FOR CONFIGURATION IN FIG. 1

| $W$ | $E_u$ | $\vec{E}_\perp$ | $H_u$ | $\vec{H}_\perp$ | $j_u$ | $j_\rho$ |
|---|---|---|---|---|---|---|
| $W \neq 0, \infty$ | $O(1)$ | $O\left(r^{-\frac{1}{2}}\right)$ | $O\left(r^{\frac{1}{2}}\right)$ | $O(\ln r)$ | $O(1)$ | $O\left(r^{\frac{1}{2}}\right)$ |
| $W = 0$ | $O\left(r^{\frac{1}{2}}\right)$ | $O\left(r^{-\frac{1}{2}}\right)$ | $O\left(r^{\frac{1}{2}}\right)$ | $O\left(r^{-\frac{1}{2}}\right)$ | $O\left(r^{-\frac{1}{2}}\right)$ | $O\left(r^{\frac{1}{2}}\right)$ |

# VI. THE JUNCTION OF RESISTIVE AND PERFECTLY CONDUCTING HALF-PLANES (FIG. 3)

The results of Table I are valid for an isolated edge of the resistive film. However, in a number of devices the film covers an aperture in a metal screen, and the model problem for such a structure is the problem of the behavior of the field in the vicinity of the common edge of resistive and perfectly conducting half-planes (Fig. 3).

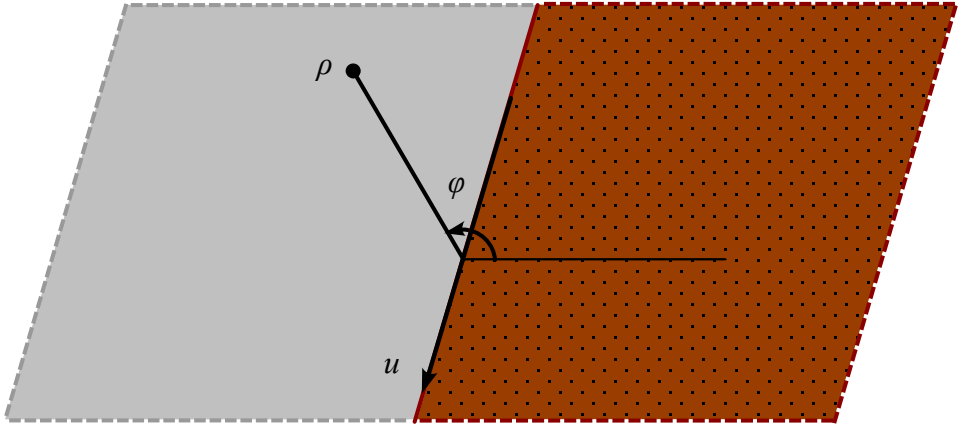

Fig. 3. Junction of resistive and perfectly conducting half-planes.

Let us consider two cases, having different symmetries of tangential magnetic field about the plane $\varphi = 0, \pi$. In the even case we immediately obtain the trivial solution (11). In the odd case the boundary conditions on the surface of the resistive film can be written as

$$E_\rho^+ = E_\rho^- = 2WH_u^+, \quad E_u^+ = E_u^- = -2WH_\rho^+. \tag{32}$$

After that it will suffice to find the field only in the "upper" half-space $0 \leq \varphi \leq \pi$. Presenting as earlier the *u*-components of the *E*- or *H*-polarized field in the form of (12) and taking account of the conditions (32) and condition $E_\rho = E_u = 0$ at $\varphi = \pi$, we shall again come to the necessity of solving the set (13) with boundary conditions

$$C_{mn}(\pi) = 0, \quad C_{mn}'(0) = -\frac{i}{2w}C_{m-1,n}(0) \tag{33}$$

for *E*-polarization or

$$C_{mn}'(\pi) = 0, \quad C_{mn}(0) = -2iwC_{m-1,n}(0) \tag{34}$$

for H-polarization. The partial solution dominating as $r \to 0$ has the form

$$E_u = U_0\sqrt{r}\left[\cos\frac{\varphi}{2} - \frac{i}{3\pi w}\cos\frac{3\varphi}{2}r\ln r + \frac{i}{3\pi w}(\varphi-\pi)\sin\frac{3\varphi}{2}r\right.$$
$$\left. - \frac{1}{30\pi^2 w^2}\cos\frac{5\varphi}{2}r^2\ln^2 r + \cdots\right],$$

$$H_\rho = -U_0\frac{i}{\zeta\sqrt{r}}\left[-\frac{1}{2}\sin\frac{\varphi}{2} + \frac{i}{2\pi w}\sin\frac{3\varphi}{2}r\ln r\right.$$
$$\left. + \frac{i}{2\pi w}\left(\frac{2}{3}\sin\frac{3\varphi}{2} + (\varphi-\pi)\cos\frac{3\varphi}{2}\right)r + \frac{1}{12\pi^2 w^2}\sin\frac{5\varphi}{2}r^2\ln^2 r + \cdots\right], \quad (35)$$

$$H_\varphi = U_0\frac{i}{\zeta\sqrt{r}}\left[\frac{1}{2}\cos\frac{\varphi}{2} - \frac{i}{2\pi w}\cos\frac{3\varphi}{2}r\ln r\right.$$
$$\left. - \frac{i}{2\pi w}\left(\frac{2}{3}\cos\frac{3\varphi}{2} - (\varphi-\pi)\sin\frac{3\varphi}{2}\right)r - \frac{1}{12\pi^2 w^2}\cos\frac{5\varphi}{2}r^2\ln^2 r + \cdots\right];$$

$$H_u = I_0\left[1 - \frac{2iw\cos\varphi}{\pi}r\ln r + \frac{2iw(\varphi-\pi)\sin\varphi}{\pi}r - \frac{w^2\cos 2\varphi}{\pi^2}r^2\ln^2 r\right.$$
$$\left. + \frac{2w^2}{\pi^2}\left(\frac{1}{2}\cos 2\varphi + (\varphi-\pi)\sin 2\varphi\right)r^2\ln r + \cdots\right],$$

$$E_\rho = i\zeta I_0\left[\frac{2iw\sin\varphi}{\pi}\ln r + \frac{2iw}{\pi}(\sin\varphi + (\varphi-\pi)\cos\varphi)\right.$$
$$\left. + \frac{2w^2\sin 2\varphi}{\pi^2}r\ln^2 r + \frac{4w^2(\varphi-\pi)\cos 2\varphi}{\pi^2}r\ln r + \cdots\right], \quad (36)$$

$$E_\varphi = -i\zeta I_0\left[-\frac{2iw\cos\varphi}{\pi}\ln r - \frac{2iw}{\pi}(\cos\varphi - (\varphi-\pi)\sin\varphi) - \frac{2w^2\cos 2\varphi}{\pi^2}r\ln^2 r\right.$$
$$\left. + \frac{4w^2(\varphi-\pi)\sin 2\varphi}{\pi^2}r\ln r + \cdots\right].$$

The main result (35), (36) is given formally in Table II, which describes the current on the resistive half-plane ($\varphi = 0$) as well as the current on the perfectly conducting half-plane ($\varphi = \pi$). From (35), (36) we conclude that the component of current parallel to the edge is discontinuous and the one perpendicular to the edge is continuous, but has a discontinuous derivative. For comparison, Table II has Meixner's result [1], which is obtained without the resistive half-plane ($W = \infty$). As in the case of an isolated half-plane, this particular case cannot be obtained from expansions (36), but requires separate consideration.

TABLE II

BEHAVIOR OF THE FIELD NEAR THE EDGE FOR CONFIGURATION IN FIG. 3

| W | $E_u$ | $\vec{E}_\perp$ | $H_u$ | $\vec{H}_\perp$ | $j_u$ | | $j_\rho$ | |
|---|---|---|---|---|---|---|---|---|
| | | | | | $\varphi = \pi$ | $\varphi = 0$ | $\varphi = \pi$ | $\varphi = 0$ |
| $W \neq 0, \infty$ | $O\left(r^{\frac{1}{2}}\right)$ | $O(\ln r)$ | $O(1)$ | $O\left(r^{-\frac{1}{2}}\right)$ | $O\left(r^{-\frac{1}{2}}\right)$ | $O\left(r^{\frac{1}{2}}\right)$ | $O(1)$ | $O(1)$ |
| $W = \infty$ | $O\left(r^{\frac{1}{2}}\right)$ | $O\left(r^{-\frac{1}{2}}\right)$ | $O\left(r^{\frac{1}{2}}\right)$ | $O\left(r^{-\frac{1}{2}}\right)$ | $O\left(r^{-\frac{1}{2}}\right)$ | — | $O\left(r^{\frac{1}{2}}\right)$ | — |

## VII. NUMERICAL RESULTS

Knowledge of the obtained behavior of the field near the edge brings substantial advantages. To illustrate these, let us use the Galerkin method to solve several integral equations for the surface current density on the resistive film in the waveguide. The general formulation of these integral equations for diffraction and dispersion problems is given in [11], [12]. Let us investigate complex problems, where the current distribution on the resistive film contains simultaneously components both tangent and normal to the edge. Vector problems of dominant mode diffraction by resistive films in rectangular waveguide were solved in papers [13]-[15]. In all of them the current density was expanded into a series containing a complete set of vector functions in each of which the component tangent (normal) to the edge remained finite (approached zero). However, the manner in which a particular component approaches zero was guessed erroneously, and, unlike the one given in Table I was assumed to be linear. After the results of the present work had been obtained we repeated the computations of [13]-[15], having changed the set of basis functions. Here we give some computed data for the problem of $TE_{10}$ mode diffraction by a resistive film in the cross section of a single mode rectangular waveguide (Fig. 4).

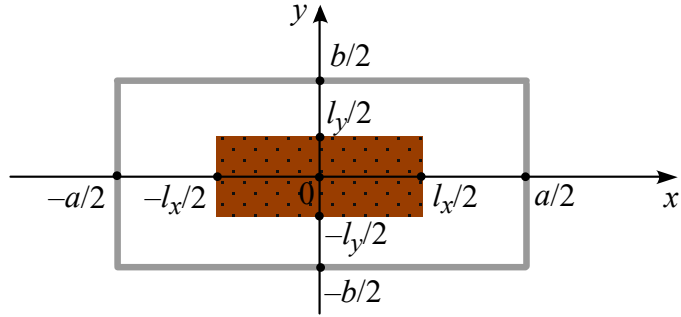

Fig. 4. Resistive film in a cross section of rectangular waveguide.

Current density components were assumed to have the form

$$j_x = \left[1-\left(\frac{2x}{l_x}\right)^2\right]^{\tau} \sum_{\mu=0}^{M^x} \sum_{\nu=0}^{N^x} A_{\mu\nu} C_{2\mu+1}^{\left(\tau+\frac{1}{2}\right)}\left(\frac{2x}{l_x}\right) C_{2\nu+1}^{\left(\frac{1}{2}\right)}\left(\frac{2y}{l_y}\right),$$

$$j_y = \left[1-\left(\frac{2y}{l_y}\right)^2\right]^{\tau} \sum_{\mu=0}^{M^y} \sum_{\nu=0}^{N^y} B_{\mu\nu} C_{2\mu}^{\left(\frac{1}{2}\right)}\left(\frac{2x}{l_x}\right) C_{2\nu}^{\left(\tau+\frac{1}{2}\right)}\left(\frac{2y}{l_y}\right),$$

(37)

where $A_{\mu\nu}, B_{\mu\nu}$ are the unknown coefficients. The value of $\tau=1$ corresponds to the set of basis functions chosen in [13] and $\tau=\frac{1}{2}$ to the real current distribution near the edge. The results of the computation of impedance $Z$ of the equivalent circuit shunt describing the analyzed discontinuity are given in Table III.

TABLE III

CONVERGENCE OF GALERKIN METOD WHILE CALCULATING THE SHUNT IMPEDANCE OF THE EQUIVALENT CIRCUIT FOR DISCONTINUITY IN FIG. 4

| N | $W=100\,\Omega$ | | $W=500\,\Omega$ | |
|---|---|---|---|---|
|   | $\tau=1$ | $\tau=\frac{1}{2}$ | $\tau=1$ | $\tau=\frac{1}{2}$ |
|   | Z | | | |
| 0 | 0.93 + i·2.49 | 0.832 + i·2.173 | 4.48 + i·2.52 | 4.041 + i·2.203 |
| 1 | 0.85 + i·2.30 | 0.833 + i·2.168 | 4.16 + i·2.37 | 4.058 + i·2.266 |
| 2 | 0.84 + i·2.24 | 0.832 + i·2.168 | 4.10 + i·2.33 | 4.059 + i·2.265 |
| 3 | 0.84 + i·2.21 | 0.832 + i·2.168 | 4.08 + i·2.30 | 4.058 + i·2.265 |
| 4 | 0.84 + i·2.20 |  | 4.08 + i·2.29 | 4.058 + i·2.265 |
| 5 | 0.83 + i·2.19 |  | 4.07 + i·2.28 |  |

The computation was made for $b=\frac{a}{2}$, $l_x=l_y=\frac{a}{4}$, $ka=\frac{4\pi}{3}$, $N^x=M^y=2$ and various $M^x=N^y=N$. It was then determined that further increase of $N^x, M^y$ does not bring about any changes in the results in Table III. It is thus clear that for any fixed $N$ at $\tau = \frac{1}{2}$ the obtained value of $Z_N$ is much more exact than at $\tau = 1$. This indicates that at $\tau = \frac{1}{2}$ the required (e.g., 1 percent) accuracy of the result is reached with a substantially lower order set than at $\tau = 1$.

Dispersion properties of waveguides with resistive films were treated in papers [16]-[18] where the manner in which the corresponding basis functions approach zero at the film edges was also chosen linear. How such a choice influences the convergence of the Galerkin method will be shown in the example of a rectangular waveguide with resistive film in its $E$-plane [16]. The cross section of the waveguide is shown in Fig. 5. (Axis $z$ is perpendicular to the plane of figure.)

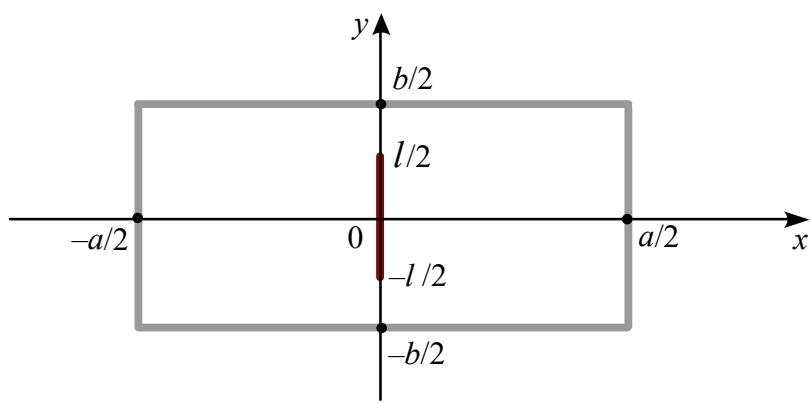

Fig. 5. Rectangular waveguide with a resistive film in $E$-plane.

Current density components on the film at $z = 0$ are sought in the form of

$$j_z = \sum_{\mu=0}^{M} A_\mu C_{2\mu+1}^{\left(\frac{1}{2}\right)}\left(\frac{2y}{l}\right),$$

$$j_y = \left[1 - \left(\frac{2y}{l}\right)^2\right]^\tau \sum_{\nu=0}^{N} B_\nu C_{2\nu}^{\left(\tau+\frac{1}{2}\right)}\left(\frac{2y}{l}\right). \tag{38}$$

The computed results for transverse wavenumber $\kappa$ for the dominant mode, which at $W \to \infty$ is transformed into the $TE_{10}$ mode of hollow waveguide, at $b = \frac{a}{2}$, $ka = 5$, $M = 4$ and various $N$ values arc given in Table IV. The numerical results of Table IV at $\tau = 1$ and $\tau = \frac{1}{2}$ for maximal $N$ differ by not more than 0.5 percent. This means, that although the results obtained in [16]-[18] were calculated at $\tau = 1$, the curves given there are correct. On the other hand, already at $N \leq 7$, knowing that the true current behavior is characterized by $\tau = \frac{1}{2}$, we can calculate $\kappa$ with an accuracy of six digits. Such high accuracy in calculating wavenumbers of complicated waveguide devices with resistive films is unprecedented.

TABLE IV

CONVERGENCE OF GALERKIN METHOD WHILE CALCULATING THE TRANSVERSE WAVENUMBER OF A RECTANGULAR WAVEGUIDE WITH A RESISTIVE FILM

| N | $\frac{l}{b}=0.8$; $W=100\,\Omega$ | | $\frac{l}{b}=0.4$; $W=500\,\Omega$ | |
|---|---|---|---|---|
|   | $\tau=1$ | $\tau=\frac{1}{2}$ | $\tau=1$ | $\tau=\frac{1}{2}$ |
|   | $\kappa a$ | | | |
| 0 | $2.636 - i \cdot 1.541$ | $2.69295 - i \cdot 1.92690$ | $2.972 - i \cdot 0.274$ | $2.95581 - i \cdot 0.31397$ |
| 1 | $2.591 - i \cdot 1.855$ | $2.58338 - i \cdot 2.01521$ | $2.957 - i \cdot 0.298$ | $2.95241 - i \cdot 0.30954$ |
| 2 | $2.578 - i \cdot 1.935$ | $2.57672 - i \cdot 2.01251$ | $2.954 - i \cdot 0.303$ | $2.95244 - i \cdot 0.30942$ |
| 3 | $2.576 - i \cdot 1.966$ | $2.57607 - i \cdot 2.01214$ | $2.953 - i \cdot 0.306$ | $2.95245 - i \cdot 0.30941$ |
| 4 | $2.574 - i \cdot 1.982$ | $2.57592 - i \cdot 2.01208$ | $2.953 - i \cdot 0.307$ | $2.95245 - i \cdot 0.30941$ |
| 5 | $2.575 - i \cdot 1.991$ | $2.57587 - i \cdot 2.01206$ | $2.953 - i \cdot 0.308$ | |
| 6 | $2.575 - i \cdot 1.996$ | $2.57585 - i \cdot 2.01205$ | $2.953 - i \cdot 0.308$ | |
| 7 | $2.575 - i \cdot 2.000$ | $2.57584 - i \cdot 2.01205$ | $2.953 - i \cdot 0.308$ | |
| 8 | $2.575 - i \cdot 2.002$ | $2.57584 - i \cdot 2.01205$ | $2.953 - i \cdot 0.309$ | |
| 9 | $2.575 - i \cdot 2.004$ | | $2.952 - i \cdot 0.309$ | |

## VIII. DISCUSSION

The use of general series (12) enabled us to describe the behavior near the edge of a resistive half-plane taken separately, (Fig. 1), as well as in conjunction with a perfectly conducting half-plane, (Fig. 3). Note that the presence of logarithmic functions in the $\psi$ expansion is a property of the field in the vicinity of the edge not only of the resistive film, but also of some dielectric wedges. Contradictions of Meixner's expansion were for the first time shown by Andersen and Solodukhov while studying the *H*-polarized field near the edge of a two-dimensional dielectric wedge, the angle of which is a rational multiple of $\pi$. However, the authors had not found the way to eliminate these contradictions. In the case of an *E*-polarized field the contradictions appear at arbitrary angles of a dielectric wedge. Makarov and Osipov [19] have shown that the series, describing the field near the edge of a dielectric wedge, should generally contain logarithmic functions. The significant feature of these series is that for arbitrary combinations of dielectric and perfectly conducting wedges logarithmic functions do not enter the first terms of the series. Presentation in the form of $E_u, H_u = \rho^\tau(C_0 + C_1\rho + o(\rho))$ is correct for all structures, considered in [1]-[4]. Thus, the first terms, determining the singularity of the field are obtained correctly in these papers. It is clear that introduction of logarithmic functions is of no principal importance while considering dielectric wedges. On the contrary, the analysis of the structures shown in Figs. 1 and 3 with no account of logarithmic functions leads to wrong results. Such an error was made by Lang in his paper [20], which addresses the resistive film geometry in Fig. 3. What the author had found was only a partial solution by defining one of the possible values of $\tau = \frac{1}{2}$. This solution does not include major terms in expansions $\vec{E}_\perp$ and $H_u$. Therefore, the author's conclusion, that the introduction of a resistive film excludes the singularity of the electric field near the edge of a perfectly conducting half-plane is erroneous. In reality, the influence of a resistive film leads to a change in the nature of the electric field singularity from algebraic to logarithmic. In conclusion, note that to date the resistive film is the only known structure whose field components have singular logarithmic behavior near the edge.